\documentstyle[prl,aps,epsfig,preprint]{revtex}
\begin{document}
\tighten

\title{One-Loop Corrections and All Order Factorization \\
         In Deeply Virtual Compton Scattering}
\author{Xiangdong Ji and Jonathan Osborne}
\bigskip

\address{
Department of Physics \\
University of Maryland \\
College Park, Maryland 20742 \\
{~}}

\date{UMD PP\#98-074 ~~~DOE/ER/40762-139~~~ January 1998}

\maketitle

\begin{abstract}
    We calculate the one-loop corrections to a general 
off-forward deeply-virtual Compton process at leading twist for both 
parton helicity-dependent and independent cases. We show that 
the infrared divergences can be factorized entirely into 
off-forward parton distributions, even when one of the
two photons is onshell. We argue that this property
persists to all orders in perturbation theory. We obtain 
the next-to-leading order Wilson 
coefficients for the general leading-twist expansion of 
the product of two electromagnetic currents in the $
\overline{\rm MS}$ scheme.

\end{abstract}
\pacs{xxxxxx}

\narrowtext

\section{introduction}

Photons, real or virtual, are known to be clean probes 
of the internal structure of the nucleon. In deep inelastic
scattering (DIS), the cross sections for absorption of 
highly-virtual photons were the first to reveal
the internal quark structure of nucleons. The parton 
distributions extracted from these cross sections 
contain important structural information 
and seriously challenge our 
understanding nonperturbative quantum
chromodynamics (QCD). Elastic 
absorption of virtual photons can be used to measure 
the electromagnetic form factors of the nucleon. 
At low virtuality, these form factors
give us direct information about the sizes and 
magnetic moments of nucleons.  At high virtuality, they are sensitive
to the leading-twist light-cone wavefunctions.
More recently, real photon elastic scattering at low 
energy has been used to extract the electromagnetic 
polarizablities of nucleons.

In a recent paper, one of us introduced deeply-virtual Compton 
scattering (DVCS) as a probe to a novel class of 
``off-forward" parton distributions (OFPD's)
\cite{ji1}. DVCS is a process in which a highly virtual
photon (with virtuality $Q^2>\!\!>\Lambda_{\rm QCD}^2$)
scatters on a nucleon target (polarized or unpolarized),
producing an exclusive final state consisting of a
high-energy real photon and a slightly recoiled 
nucleon. With the virtual photon in the Bjorken limit, 
a QCD analysis shows that the scattering
is dominated by the simple mechanism in which a quark
(antiquark) in the initial nucleon absorbs the virtual photon, 
immediately radiates a real one, and falls back to 
form the recoiled nucleon. 

Several interesting theoretical papers have 
since appeared in the literature, which studied
the DVCS process further. In Ref. \cite{ra1}, 
the single-quark scattering was recalculated 
using a different, but equivalent definition of 
the parton distributions.  The evolution equations of the distributions
were derived and some general aspects of 
factorization were discussed.
In Ref. \cite{ji2}, the evolution equations for
OFPD's were derived and the leading-twist DVCS cross sections 
were calculated at order $\alpha_s^0$. Some past and recent 
studies of OFPD's can be found in \cite{ofpd}. 
In Ref. \cite{guichon}, estimates of these cross sections were made 
at COMPASS and TJNAF energies. In Ref. \cite{chen}, 
the DVCS process was considered as a limit of unequal mass
Compton scattering, which was studied from the point of view 
of the operator product expansion. Some early studies 
of unequal mass Compton processes can be found 
in Refs. \cite{wana,muller1}. In Ref. \cite{di}, 
a number of suggestions were made to test the leading twist
dominance in DVCS at finite $Q^2$. In a Rapid Communication 
paper, the present authors studied ${\cal O}(\alpha_s)$ corrections
to DVCS for the parton helicity-independent case\cite{jon}. 
In Refs. \cite{muller,man}, the same issue was investigated
from different perspectives. The present paper is an expanded 
presentation of our results in Ref. \cite{jon}. 

The main motivation for the present study is to see
if the theoretical basis for the DVCS process is up to par 
with other well-known perturbative QCD processes.
More explicitly, we discuss the existence of a factorization 
theorem for this process. For general two
virtual photon processes in the Bjorken limit, the 
factorizability is suggested by studies of deep
inelastic scattering. In the case of DVCS, where one of the 
photons is onshell, the situation could be different. 
Potential infrared problems can arise 
because of the additional light-like vector 
in this special kinematic limit. However, it is believed 
that these complications will 
not ruin the factorization properties \cite{ji2}.

To see factorization at work, it is instructive
to work out one-loop examples. We 
will do this explicitly in section III. For consistency, 
we consider the unphysical process
of DVCS on onshell quark and gluon ``targets.'' 
To ensure gauge invariance, we regularize
the infrared divergences by going to $d=4+\epsilon$
dimensions. For completeness we have considered both
the symmetric and antisymmetric parts of the amplitudes, 
which are related to helicity-independent and dependent
parton distributions, respectively. The only omission
is the gluon helicity flip amplitude, which 
will be discussed in Ref. \cite{hoodbhoy}. As expected,
our result contains collinear infrared divergences 
which can be interpreted as the one-loop 
perturbative parton distributions, as we will show in Section 
IV. This property is independent of the 
special kinematic limit of DVCS. 

A general proof of the DVCS factorization 
was first given by Radyushkin in his approach
based on $\alpha$-representation \cite{ra1}.  In this
paper, we give an alternative proof using 
the tools developed by Libby, Sterman, Collins  
and others \cite{factor}. According to these, 
one can represent the infrared sensitive contributions 
in a generic Feynman diagram with reduced diagrams. 
These reduced diagrams have intuitive physical 
significance and are easy to identify. General 
power counting rules can be used to select
leading reduced diagrams in a process. A recent application 
of the method can be found in Ref.\cite{collins}.  We show
in Section V that the leading reduced diagrams for DVCS
do not contain any soft divergences and are in fact 
exactly the same as those present when the final state photon 
is deeply virtual. The collinear
divergences in the reduced diagrams can be attributed  
to those of OFPD's when calculated in perturbation 
theory. Therefore we conclude that factorization
for DVCS is in the same footing as other 
well-known examples like deep inelastic scattering. 

The factorization property of the general two 
virtual photon process can be summarized beautifully
in terms of Wilson's operator product expansion. 
This expansion requires operators with total derivatives
\cite{chen,wana,muller1} to describe the 
off-forward nature of the process. It is well-known that
these derivative operators contribute to the 
wavefunctions of mesons \cite{erbl}.  
In section VI, we convert our one-loop results 
into Wilson coefficients of the twist-two 
operators in the $\overline{\rm MS}$ scheme. 
Together with the two-loop anomolous dimensions
of these operators, they provide the necessary 
ingredients for calculating 
DVCS at the next-to-leading order.

We summarize and discuss our results in Section VII.

\section{Kinematics and Parton distributions}

Although our ultimate interest is in deeply virtual Compton
scattering, we start by  considering a general Compton
process involving two offshell photons with different 
virtualities.  This and a suitable choice of kinematic 
variables allows us to exploit the full symmetry of 
the problem.  In the general Compton process, a virtual photon 
of momentum $q+\Delta/2$ is absorbed by a hadron of 
momentum $P-\Delta/ 2$, which then emitts a virtual photon with momentum 
$q-\Delta/ 2$ and recoils with momentum $P+{\Delta /2}$.  
The three independent external momenta can be expanded in 
terms of the light-cone vectors 
\begin{eqnarray}
       p^\mu &=& \left(p^+,0,0,p^+\right) \ ,  \\
      n^\mu &=& {1\over {2 p^+}}\left(1,0,0,-1\right) \ , \nonumber
\end{eqnarray}
where the 3-direction is chosen as the direction of the average hadron 
momentum ($P$), and two transverse vectors.  
In an expansion, we call the coefficient of 
$p^\mu$ the + component and that of $n^\mu$ the $-$ component. 
Thus we write 
\begin{eqnarray}
      P^\mu &=& p^\mu+{M^2-{t/4}\over 2} n^\mu \ ,  \nonumber\\
      q^\mu &=& -\zeta  p^\mu+{Q^2\over {2 \zeta}}n^\mu\ ,  \\
     \Delta^\mu &=&-2\xi p^\mu+\xi(M^2-t/4)n^\mu+\Delta_T^\mu \ , \nonumber
\end{eqnarray}
where $M$ is the hadron mass (which is taken to be the same for the
initial and final hadrons), $t=\Delta^2$, $Q^2$ is the virtuality of
$q^\mu$, $\xi$ is a measure of the difference of the virtualities of 
the two external photons, $\zeta$ is defined as   
\begin{equation}
\zeta={Q^2\over{2x_B(M^2-t/4)}}\left(-1+\sqrt{1+{4x_B^2(M^2-t/4)
\over Q^2}}\right),
\end{equation}
and $\Delta_T^\mu$ is a vector in the transverse directions which has 
squared length $-t\left(1-{\xi^2}\right)-4\xi^2 M^2$.  We have also
introduced $x_B={Q^2/(2 P\cdot q)}$, the analogue of the Bjorken
scaling variable in this off-forward process.  We  note that these 
expressions limit the range of $\xi$ to
\begin{equation}
     \xi^2 \leq {-t\over -t+4M^2}\ ,   
\end{equation}
for  fixed $t$, or the range of $t$ to 
\begin{equation}
     -t\geq {4\xi^2M^2\over 1-\xi^2}\ , 
\end{equation}
for  fixed $\xi$.      

In the Bjorken limit, these expressions simplify considerably. Since we 
consider only the leading twist in this paper, we may neglect all but the 
$+$ components of $P^\mu$ and $\Delta^\mu$ (in order to form large 
scalars, one must dot the $+$ component of 
a vector with the $-$ component of $q$).
Hence, in the limit $Q^2\rightarrow\infty$ ($t$ remaining finite), we may write
\begin{eqnarray}
     && P^\mu \sim p^\mu \ ,  \nonumber\\
     && q^\mu\sim -x_B p^\mu+{Q^2\over 2 x_B} n^\mu \ , \\
     && \Delta^\mu\sim -2\xi p^\mu \ .  \nonumber
\end{eqnarray}     
Here, we note that the external invariants have been reduced from six to 
three by enforcing kinematics and taking the Bjorken limit.  We express these 
three scalars in terms of one mass scale, $Q^2$, and two dimensionless 
parameters, $x_B$ and $\xi$.  When we introduce the parton
distributions, our expressions will also involve the parton 
light-cone momentum variable $x$. Hence, the final result
will be expressed as algebraic functions of $x_B$, $x$, and $\xi$ 
multiplied by the appropriate power of $Q^2$.

\begin{figure}
\label{fig1}
\epsfig{figure=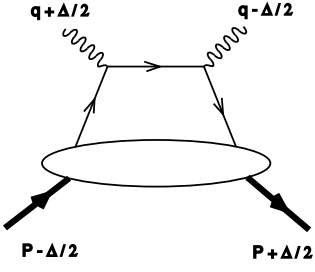,height=3.5cm}
\caption{The handbag diagram for the general two 
photon process.}
\end{figure}     

Our goal is to factorize the short and long distance 
physics of the Compton amplitude in the Bjorken limit, 
\begin{equation}
   T^{\mu\nu} = i\int d^{\, 4}z\,
     e^{\,i q\cdot z}\left\langle P+{\Delta\over 2}
      \left|TJ^\nu\left({z\over 2}
           \right) J^\mu \left(-{z\over 2}\right)
       \right|P-{\Delta\over 2}\right\rangle \ ,
\label{comam} 
\end{equation}
where $J^\mu = \sum_q e_q \bar{\psi_q}
\gamma^\mu\psi_q$ is the electromagnetic
current and $\psi_q$ is the bare quark field of flavor $q$ and
charge $e_q$. The simplest Feynman diagram for this process
is shown in Fig. 1, where a quark comes out of the nucleon
blob, scatters, and rejoins the nucleon blob. While 
the scattering involves a large momentum transfer and can
be calculated in perturbation theory, the nucleon blob
with two quark legs is related to the baryon structure 
and is nonperturbative. For more complicated graphs, as 
will be discussed throughout this paper, the Compton amplitude
can be separated analogously into soft and hard contributions.
In the remainder of this section, we will highlight some
important aspects of the soft part.

The nonperturbative contribution to the Compton amplitude
in Eq. (\ref{comam}) can be expressed in terms of
off-forward parton distributions contained in the 
parton density matrices\cite{ji2}. For quarks, 
we call the density matrix $M_{\alpha \beta}$, where $\alpha$ 
and $\beta$ are Dirac indices, and expand it in 
terms of the Dirac matrices.  At leading twist, 
$M_{\alpha\beta}$ is just the light-cone correlation function,  
\begin{eqnarray}
     M^q_{\alpha \beta}\left(x,\xi\right) &=& \int 
{d\lambda \over 2\pi} e^{-i\lambda x} \left\langle 
P+{\Delta\over 2}\left|\bar\psi^q_\beta 
\left({\lambda\over 2} n\right){\rm P}
\left\lbrace e^{-ig\int_{-{\lambda\over 2}}^{\lambda\over 2} n 
\cdot A\left( \zeta n \right) d\zeta}\right\rbrace\psi^q_\alpha
\left(-{\lambda\over 2} n\right)
\right|P-{\Delta\over 2}\right\rangle \nonumber  \\
 &=& {1\over 2} F_q {\not\! p}_{\alpha\beta} 
+ {1\over 2} \tilde F_q \left(\gamma_5\not\! 
p\right)_{\alpha\beta}+\cdots,  
\end{eqnarray}
where the ellipses denote contributions  
either of higher twist or chiral-odd structure, 
which do not contribute to the leading 
process under consideration.  The P symbol denotes the path ordering 
of the exponential, which makes this 
expression gauge invariant.  It is necessary to include this gauge link 
whenever one is not working in the 
light-cone gauge ($A^+=0$). 
Multiplying by $\not\! n_{\beta\alpha}$ and 
$(\not\! n\gamma_5)_{\beta\alpha}$ and taking traces,  
we project out the same distributions as considered 
in \cite{ji2}:
\begin{eqnarray}
      F_q(x,\xi) &=&{1\over 2} \int {d\lambda \over2\pi}
          e^{-i\lambda x}\left\langle P+{\Delta\over 2}\left|
\bar \psi_q\left({\lambda \over 2} n\right)
  {\rm P}\{\}\not\! n \psi_q\left(-{\lambda \over 2} n\right) 
\right|P-{\Delta\over 2}\right\rangle \ , \\  
     \tilde F_q(x,\xi) 
    & =&{1\over 2} \int {d\lambda \over 2\pi}
          e^{-i\lambda x}\left\langle P+{\Delta\over 2}\left|
\bar \psi_q\left({\lambda \over 2} n\right)
   {\rm P}\{\}     \not\! n\gamma_5 \psi_q\left(-{\lambda \over 2} n\right) 
\right|P-{\Delta\over 2}\right\rangle \ .
\end{eqnarray}
We have suppressed the renormalization 
scale $\mu$ which is always present in defining a 
parton distribution. We have also suppressed the $t$
dependence because it will not affect most of the discussions
in this paper.
     
At next to leading order, gluons also contribute to 
the Compton process. Although it is nontrivial to show, the
twist-two gluon distributions are contained in the following
gauge-invariant light-cone correlations ($\epsilon^{0123}=+1$), 
\begin{eqnarray}
      G^{\mu\nu\alpha\beta}(x,\xi)&=&\int{d\lambda\over 
2\pi}e^{-i\lambda x}\left\langle P+{\Delta\over 2}\left|
 F_a^{\mu\nu}\left({\lambda\over 2}n\right){\rm P}\left\lbrace 
e^{-ig\int_{-{\lambda\over 2}}^{\lambda\over 2}n\cdot 
A(\zeta n)d\zeta}\right\rbrace_{
ab} F_b^{\alpha\beta}\left(-{\lambda\over 2}n\right)
\right|P-{\Delta\over 2}\right\rangle\nonumber \\
&=& -xF_G(x,\xi)\left(g^{\mu\alpha}p^\nu 
p^\beta -g^{\mu\beta}p^\nu p^\alpha +g^{\nu\beta}
p^\mu p^\alpha -g^{\nu\alpha}p^\mu p^\beta\right) \\ 
&&+ix\tilde F_G(x,\xi)\left(\epsilon^{\mu\alpha\gamma\delta}
p^\nu p^\beta -\epsilon^{\mu\beta\gamma\delta}
p^\nu p^\alpha+\epsilon^{\nu\beta\gamma\delta}
p^\mu p^\alpha -\epsilon^{\nu\alpha\gamma
\delta}p^\mu p^\beta\right)n_\gamma p_\delta +\cdots , \nonumber
\end{eqnarray}
where the ellipses denote higher twist contributions and
an additional twist-two term which involves gluon helicity
flip and will not be considered in this paper \cite{hoodbhoy}. 
Again, $\rm P\lbrace\rbrace$ denotes path ordering 
(we note that here the gauge link is in the adjoint 
representation of $SU(3)$).  
The off-forward gluon distribution 
functions $F_G$ and $\tilde F_G$ may be isolated by contraction and are
\begin{eqnarray}
      F_G=-{1\over 2x}\int {d\lambda\over 2\pi}e^{-i\lambda x}
\left\langle P+{\Delta\over 2}\left| F_a^{\mu\alpha}
\left({\lambda\over 2} n\right){\rm P}\left\lbrace\right\rbrace_{ab} 
F_{b\,\,\alpha}^\nu\left(-{\lambda\over 2}n\right)\right|
P-{\Delta\over 2}\right\rangle n_\mu n_\nu\, \ , \\
    \tilde F_G=-{i\over 2x}\int{d\lambda\over 2\pi}
e^{-i\lambda x}\left\langle P+{\Delta\over 2}\left| 
F_a^{\mu\alpha}\left({\lambda\over 2}n\right) 
{\rm P}\left\lbrace\right\rbrace_{ab}\tilde F_{b\,\,\alpha}^\nu
\left(-{\lambda\over 2}n\right)\right|P-{\Delta
\over 2}\right\rangle n_\mu n_\nu \ .
\end{eqnarray}
Here, we have defined the dual field strength 
tensor $\tilde F^{\mu\nu}={1\over 2}
\epsilon^{\mu\nu\alpha\beta}F_{\alpha\beta}$.
     
It is easiest to see the connection 
of the above gluon distributions with the nonperturbative 
structure arising from Feynman diagrams in the 
light-cone gauge. In this gauge, the gauge link is 
just the unit operator in the adjoint representation  
and field strength tensors with one + index $F^{+\mu}$ simplify
to $\partial^+A^\mu$. Fourier transformation to 
momentum space yields
\begin{eqnarray}
      F_G  &=&-{x_+ x_-\over 2x}{1\over \rm VT}\int{d^4\ell\over
{(2\pi)}^4}\delta (x-\ell\cdot n) \nonumber \\ && \times \left\langle 
    P+{\Delta\over 2}\left|TA^\mu_a\left(\ell+{\Delta\over 2}\right) 
A^\nu_a\left(\ell-{\Delta\over 2}\right)\right|P-{\Delta\over 2}
   \right\rangle g^\perp_{\mu\nu} \ ,  \\
    \tilde F_G  &=& -i{x_+x_-\over 2x}{1\over \rm VT}
       \int{d^4\ell\over {(2\pi)}^4}\delta (x-\ell\cdot n) \nonumber \\
 && \times \left\langle P+{\Delta\over 2}\left|TA^\mu_a
\left(\ell+{\Delta\over 2}\right)A^\nu_a
\left(\ell-{\Delta\over 2}\right)\right|P-{\Delta\over 2}
\right\rangle \epsilon_{+-\mu\nu}\ ,
\end{eqnarray}
where we have defined $x_+ = x+\xi$ and $x_- = x-\xi$.  
VT represents $(2\pi)^4\delta^4(0)$, the space-time 
volume of our system. 
In a factorized calculation
of the Compton amplitude involving gluons, the gluonic 
indices in the hard part will be contracted with 
the tensor
\begin{eqnarray}
   &&  {1\over \rm VT}\int{d^4\ell\over
{(2\pi)}^4}\delta (x-\ell\cdot n)  \times \left\langle
    P+{\Delta\over 2}\left|TA^i_a\left(\ell+{\Delta\over 2}\right)
A^j_a\left(\ell-{\Delta\over 2}\right)\right|P-{\Delta\over 2}
   \right\rangle \nonumber \\
  && = -{x\over x_+x_-}\left(F_G(x)g^{ij} + i\tilde
F_G(x)\epsilon^{+-ij}\right) \ .
\label{gluedist} 
\end{eqnarray}

In the above definitions, we have assumed that we are working
in 3+1 space-time dimensions. However, to regularize the
ultraviolet and infrared divergences arising from loop
diagrams, it is convenient to generalize them to 
$d$ dimensions. Let us first consider the quark density matrix 
in Eq. (8). Because the spinors are kept in 4 dimensions,  
the first term on the right hand side generalizes to $d$ 
dimensions without change. The second term, however, involves  
$\gamma_5$ which has no unique extension. Different choices, 
in the end, define different factorization schemes. If one 
uses the t' Hooft-Veltman definition ($\gamma_5=i\gamma^0\gamma^1
\gamma^2\gamma^3$)\cite{veltman}, one usually introduces an extra 
renormalization constant $Z_5$ so that the non-singlet 
axial currents are conserved. An alternative choice
is offered by Chanowitz, Furman, and 
Hinchcliffe \cite{chanowitz}, which employs
the usual four-dimensional rules 
\begin{eqnarray}
      \lbrace{\gamma^\mu,\gamma_5\rbrace} &=& 0 ~~~~~~~~~~~~~~~~~~~~
\forall \mu\in\lbrack 0,d\rbrack \nonumber  \\
      {\rm Tr}\lbrack \gamma_5\gamma^\alpha\gamma^\beta\gamma^\gamma
\gamma^\delta \rbrack &=& -(4+{\cal O}(\epsilon))\,\,
i\epsilon^{\alpha\beta\gamma\delta} \ . 
\label{tr}
\end{eqnarray}
The ambiguity in the second equation
does not affect calculations 
as long as there are no anomalies in the problem.
In the case that there is an anomaly, the ambiguity can be 
fixed by imposing the relevant 
Ward identities.

We now turn to the gluon density matrix in Eq. (\ref{gluedist}). 
$F_G(x)$ contains an average over gluon polarizations.  To make this  
consistent with the number of transverse polarization states
available to gluons in $d$ dimensions, we 
multiply this term by $1/(1+\epsilon/2)$. The polarized gluon
density is related to the antisymmetric combination of the gluon fields 
$F^{+1}$ and $F^{+2}$. This does not change after going to
$d$ dimensions if the target polarization is kept
the same. Hence, we have left that term as it is. 

\section{One-loop compton amplitudes \\ on quark and gluon ``targets''}     

In this section, we present a one-loop calculation of the 
general Compton scattering on onshell quark and gluon
``targets'' in the Bjorken limit. The result will be used in 
the next section to show that the factorization of soft
and hard contributions can be done consistently with
the definition of off-forward parton distributions. In particular, 
this property does not change in the limit of 
a real final state photon. 
Our result will also be used to derive a generalized 
operator product expansion to next-to-leading order.  
For the convenience of the reader, we are going to 
spell out some technical details of the one loop calculation.  
We believe that some of the techniques, like 
the cancellation of propagators and  
light-front coordinate integration, will be useful
in other contexts. 

We begin with an onshell quark ``target''.  
Here, there are two diagrams at 
leading order (LO) and eight at next-to-leading order 
(NLO).  Half of these diagrams are shown in Fig. 2. The
other half will be taken into account by using the 
crossing symmetry, i.e., the simultaneous replacement of $q\rightarrow -q$
and $\mu\leftrightarrow\nu$. The terms with 
$x_B\rightarrow -x_B$ in the following formulas reflect this 
contribution. Because of time reversal invariance,
the Compton amplitude is also an even function of $\xi$, i.e., 
symmetric under $\xi\rightarrow -\xi$.   
This symmetry relates the left and right 
vertex diagrams (Figs. 2c and 2d, respectively) to each other.  
On the other hand, the quark self-energy diagram and the
box diagram are themselves $\xi$-symmetric. 
This symmetry not only 
allows us to reduce the number of graphs at NLO from four to three, but also 
becomes a powerful tool which helps us compute each amplitude, as we  
illustrate later.

\begin{figure}
\label{fig2}
\epsfig{figure=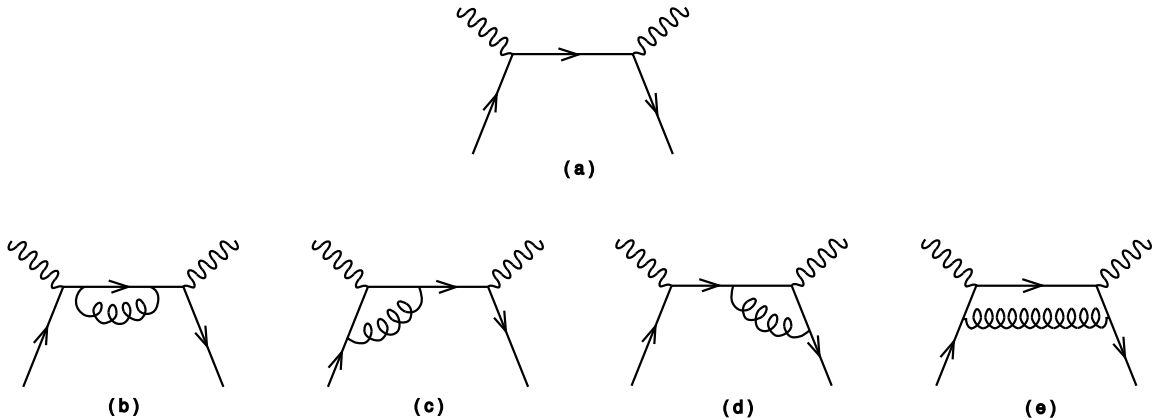,height=5.5cm}
\caption{Diagrams for Compton scattering on an onshell quark
to order $\alpha_s$.}
\end{figure}           
Before presenting the details of our calculation, it is necessary to 
discuss the issues of onshell reduction and ultraviolet 
divergences. We calculate diagrams in $d$ dimensions
and use Feynman's gauge. 
As such, we can take the quark target to be massless and 
take the onshell limit at the beginning of our calculation. 
In the modified minimal subtraction scheme ($\overline{\rm MS}$), 
the renormalized single-quark propagator has a residue, $\delta Z$,  at the
pole ${\not\! p} =0$. According to Lehmann-Symanzik-Zimmerman
reduction formula, we calculate
the onshell physical matrix element of an operator $\hat {\cal O}_R$
as
\begin{equation}
        \langle p |\hat {\cal O}_R |p \rangle 
     = \lim_{{\not p}\,\rightarrow 0}
       {\delta Z}\sum_{G} G_{\cal O}\,\,\,\, ,
\end{equation}
where $G_{\cal O}$ is the set of all amputated connected 
graphs with one insertion of $\hat {\cal O}_R$  
in renormalized perturbation theory.
The factor $\delta Z$ is infrared divergent for massless quarks and 
equals $Z_F^{-1}$ in the present calculational scheme. Since 
$J^\mu = Z_F \bar \psi_R \gamma^\mu \psi_R$, all renormalization
constants, including the subtraction for the quark self-energy,
cancel at one-loop level.  Therefore, $T^{\mu\nu}$ for single quark and
gluon ``targets" can be calculated just from 
the graphs shown in Fig. 2. 

  Examining these graphs, we see that the self-energy diagram 
contains a loop integral with two Feynman denominators, the vertex 
diagram contains one with three, and the box with four. 
A one-loop integral with two propagators is straightforward.  
Difficulties arise, however, with the calculation of three and 
especially four-propagator integrals.  These difficulties 
may be avoided in this 
calculation because of several simplifications.  
Consider first the box diagram.  The loop integral 
is of the form (with momentum routing as shown in Fig. 3)
\begin{equation}
      \int {d^d k\over {(2\pi)}^d} {{{\rm Tr}
\lbrack\gamma^\alpha\left({\not\! k}
-\xi\not\! p \right)\gamma^\nu\left({\not\! k}+{\not\!q}\right)
\gamma^{\mu}\left({\not\! k}+\xi\not\! p \right)\gamma_\alpha
{\not\! p}(\gamma_5)\rbrack}\over
 {(k+\xi p )^2(k-\xi p)^2(k-p)^2(k+q)^2}}\ , 
\end{equation}
where we have replaced $\Delta^\mu$ by $-2\xi p^\mu$.  
In order to simplify the integral, 
we express the trace as a sum of terms which
cancel one of the propagators. This can be done because
both $k^2$ and  $2p\cdot k$ can be written as 
linear combinations of $(k+\xi p)^2$
and $(k-\xi p)^2$, and the trace vanishes 
whenever $k^-$ and $k_T^2$ do.  

\begin{figure}
\label{fig3}
\epsfig{figure=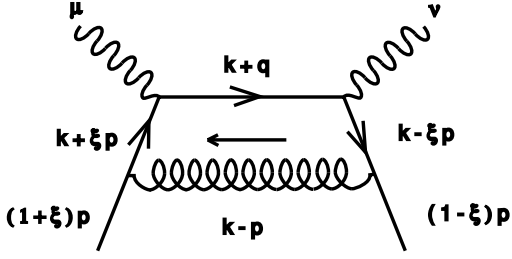,height=3.5cm}
\caption{The momentum flow in the box diagram}
\end{figure}                
Now that we have shown that a denominator can be cancelled, we have
effectively reduced the four-propagator problem to a three-propagator 
one.  Since we have only shown that the numerator will be a 
linear combination of two different denominators, rather that proportional 
to one, the four-propagator integral will in general become 
two three-propagator integrals.  However, we may use 
the $\xi$ symmetry by writing
\begin{eqnarray}
      k^2={1\over 2}&&\left(k+\xi p\right)^2+(\xi\rightarrow 
-\xi) \ , \nonumber\\
      2p\cdot k={1\over 2\xi}&&\left(k+\xi p \right)^2+
(\xi\rightarrow -\xi) \ .
\end{eqnarray}
     In this way, we consider only the $(k+\xi p )^2$ cancellation
and let the symmetry take care of the rest.  We note that the integral we 
now have is  exactly the same (up to numerator differences) as that 
arising from the right vertex correction.
We will see that this basic integral is the only one we must calculate to 
obtain both the polarized and unpolarized amplitudes for both the quark and 
gluon contributions (if we forget, for the moment, the simple self-energy
 diagram).  The integral has the form
\begin{equation}
      \int {d^dk\over{(2\pi)}^d} {Numerator\over{[(k-p)^2+i\epsilon]
[(k-{\xi}p)^2+i\epsilon][(k+q)^2+i\epsilon]}}\,\,.
\label{int}
\end{equation}
We have found that this integral is easily done 
in light-cone coordinates by expanding $k^\mu$ 
in terms of
the light cone momenta, $p^\mu$ and $n^\mu$.  We first 
do the $k^-$ integration by contour in the unphysical 
region of large $x_B$, and then the transverse integrations. The
$k^+$ integration is left until the end.  If we write $k^+=yp^+$, 
the value of the integral (\ref{int}) is 
\begin{equation}
       {i\over 16\pi^2}\left({Q^2\over 4\pi}\right)^{{\epsilon\over 
2}}{\Gamma(-{\epsilon\over 2})\over 1-\xi }\left\lbrace\int_a^{x_B}
dy\left({{y-a}\over{x_B-a}}\right)^{1+{\epsilon\over 2}}\left(
1-{y\over x_B}\right)^{{\epsilon\over 2}}N\Big|^{a=1}_{a={\xi}}\right\rbrace,
\end{equation}
where we have defined $N$ according to
\begin{eqnarray}
      Numerator\Big|_{k^-={k_T^2\over 2p^+(y-a)}}=\alpha+\beta k_T^2; 
\nonumber\\
      N =\beta-{\alpha\over Q^2}{{x_B(x_B-a)}\over{(y-a)(x_B-y)}}.
\end{eqnarray}
     Doing the $y$-integrals requires some care because a delicate 
cancellation
must occur if one is to get finite result, but the treatment is 
straightforward.  After the $\xi$ and crossing symmetries are used, 
we find the full NLO result for the symmetric quark amplitude
\begin{eqnarray}
      T^{(ij)}_q &=& -g^{ij}\sum_{q'} e_{q'}^2 \delta_{qq'}\left\lbrace
{1\over x_B-1}-{\alpha_s C_F\over 4\pi}\left({Q^2e^{\gamma_E}\over
4\pi\mu^2}\right)^{{\epsilon\over 2}}\left\lbrace {3\over
{x_B-1}}\left(-{2\over\epsilon}+3\right)\right.\right.\nonumber\\
&& -{1\over\xi}\left\lbrack\left({2\xi\over x_B^2-1}
+{x_B+\xi\over 1-\xi^2}\right)
\left(-{4\over\epsilon}+3-\ln\,\left(1-{\xi\over x_B}\right)
\right)-3\,{x_B-\xi\over 1-\xi^2}\right\rbrack \ln\left(1-{\xi\over
x_B}\right)\nonumber \\ && \left. 
+\left\lbrack  \left({x_B+1\over 1-\xi^2}+{2\over
x_B-1}\right)  \left(-{4\over\epsilon}+3-\ln\, \left(1-{1\over
x_B}\right)\right)-3\,{x_B-1\over 1-\xi^2} \right.\right. \\
\nonumber \\ && \left.\left.\left. - {3\over x_B-1}\right
\rbrack \ln\left(1-{1\over x_B}\right)\right\rbrace 
+(x_B\rightarrow -x_B)\right\rbrace\nonumber
\end{eqnarray}                                 
and the antisymmetric amplitude
\begin{eqnarray}
  T^{[ij]}_q &=&i\epsilon^{\alpha\beta ij}n_\alpha p_\beta \sum_{q'} 
    e_{q'}^2 \delta_{qq'} \left\lbrace {1\over x_B-1}-{\alpha_s C_F\over
4\pi}\left({Q^2e^{\gamma_E}\over 
4\pi\mu^2}\right)^{{\epsilon\over 2}}\left\lbrace {3\over
x_B-1}\left(-{2\over\epsilon}+3\right)\right.\right.\nonumber\\
&& \left.\left.-\left\lbrack\left({2x_B\over x_B^2-1}+{x_B+\xi\over
1-\xi^2}\right)\left(-{4\over\epsilon}+3-\ln\,\left(1-{\xi\over
x_B}\right)\right)-{x_B-\xi\over 1-\xi^2}
\right\rbrack \ln\left(1-{\xi\over
x_B}\right)\right.\right.  \\
&& +\left\lbrack\left({x_B+1\over 1-\xi^2}+{2\over
x_B-1}\right)\left(-{4\over\epsilon}+3-\ln\,
\left(1-{1\over x_B}\right)\right)-{x_B-1\over
1-\xi^2} \right.\nonumber \\ && \left.\left.\left.
-{3\over x_B-1}\right\rbrack \ln\left(1-{1\over
x_B}\right)\right\rbrace-(x_B\rightarrow -x_B)\right\rbrace\nonumber \ . 
\end{eqnarray}   
Here we have introduced $C_F={\rm Tr}[t_at_a]=
{N^2_c-1\over 2N_c}$ in $SU(N_c)$, where $N_c$ is the 
number of colors.  We note that the divergences in 
these amplitudes are, in fact, infrared divergences since renormalization
 has removed all ultraviolet singularities.  Their 
presence signals the existence of nonperturbative physics 
in the process. As mentioned earlier, these divergences will be 
factorized into nonperturbative matrix elements whose values can 
be extracted from experiment.  We will explicitly show this in
the next section.  For now, we summarize the results of the gluon 
piece of the calculation.

\begin{figure}
\label{fig4}
\epsfig{figure = 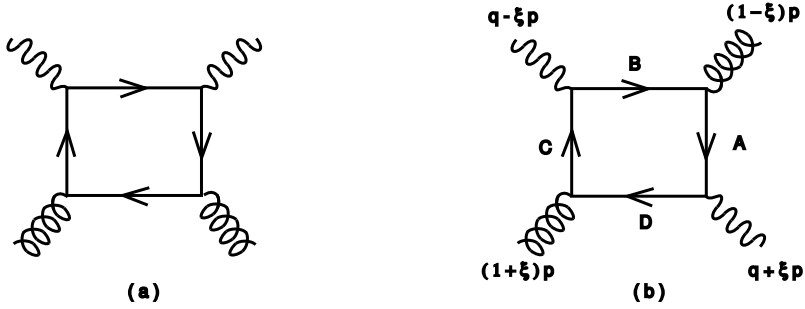, height=4cm}
\caption{Diagrams for gluon Compton scattering at one loop.}
\end{figure}           
There are six graphs which contribute to the LO amplitude for gluon-photon
scattering.  These six can be reduced to three by reversing the 
fermion number
flow, and one of these can be eliminated by crossing symmetry.  
The two distinct graphs we must calculate are shown in Fig. 4.  
The denominator of Fig. 4a is
identical to that of the quark box.  Again, the numerator is seen 
to vanish 
whenever $k^2$ and $2p\cdot k$ do, which allows us to cancel one of 
the propagators exactly as above.  Fig. 4b is somewhat more tricky.  
This diagram is 
itself symmetric under both crossing and $\xi$ symmetry.  Labeling 
the momenta as shown, we see that under an integral the symmetry 
$q\rightarrow -q$ is 
equivalent to $A\leftrightarrow B$ and $C\leftrightarrow D$ and 
$\xi\rightarrow -\xi$ is equivalent to $B\leftrightarrow D$ and 
$A\leftrightarrow C$.  We also note that here it 
is possible to represent 1 as a linear combination of Feynman  
denominators, which guarantees our ability to cancel one.  
Of course, since 1 does not depend on $q$ or $\xi$, it may be represented 
in the symmetric form
\begin{equation}
      1=-{1\over 2(1-\xi)p\cdot q}A^2+(\xi\rightarrow -\xi)+(q\rightarrow -q).
\end{equation}
    Now it remains only to calculate the trace in a symmetric way and 
substitute the result into the formulae of the quark calculation.  
Averaging the gluon polarization for the symmetric
amplitude in $d$-dimensions, one finds  
\begin{eqnarray}
       T_g^{(ij)}&=&-{\alpha_sT_F\over 2\pi}g^{ij}\left(\sum_{q}
e_q^2\right)\left({Q^2e^{\gamma_E}\over 4\pi\mu^2}\right)^{{\epsilon\over 2}}
{1\over 1-\xi^2}
\nonumber\\ &&
\left\lbrace\left\lbrack\left(1+2x_B{x_B-1\over 1-\xi^2}\right)\left(
-{4\over\epsilon}+4-\ln\,\left(1-{1\over x_B}\right)\right)-2\right
\rbrack \ln\left(1-{1\over x_B}\right)\right.\\ &&
-\left.2x_B\left\lbrack\left({x_B\over 1-\xi^2}-{1\over 2\xi}\right)
\left(-{4\over\epsilon}+4-\ln\,\left(1-{\xi\over x_B}\right)\right)+
{1\over\xi}\right\rbrack\left(1-{\xi\over x_B}\right)\ln\left(1-
{\xi\over x_B}\right)\right.\nonumber\\
&&\left. +(x_B\rightarrow -x_B)\right\rbrace\nonumber \ , 
\end{eqnarray}  
for the symmetric amplitude and 
\begin{eqnarray}
      T_g^{[ij]}&=&i{\alpha_sT_F\over 2\pi}\epsilon^{\alpha\beta ij}
n_\alpha p_\beta\left(\sum_{q}e_q^2\right)\left({Q^2e^{\gamma_E}\over 4
\pi\mu^2}\right)^{{\epsilon\over 2}}{1\over 1-\xi^2} \nonumber\\
&&\left\lbrace\left\lbrack\left(1+2{x_B-1\over 1-\xi^2}\right)\left(
-{4\over\epsilon}+4-\ln\,\left(1-{1\over x_B}\right)\right)-2\right
\rbrack \ln\left(1-{1\over x_B}\right)\right.\\ &&
-\left.2x_B\left\lbrack{1\over 1-\xi^2}\left(-{4\over\epsilon}+4-\ln\,
\left(1-{\xi\over x_B}\right)\right)\right\rbrack\left(1-{\xi\over
 x_B}\right)\ln\left(1-{\xi\over x_B}\right)\right.\nonumber\\
&& -\left. (x_B\rightarrow -x_B)\right\rbrace\ \ ,\nonumber
\end{eqnarray}
for the antisymmetric amplitude.  
We have defined $T_F= {\rm Tr}[t_at_b]=
{1\over 2}\delta_{ab}$.

\section{one-loop factorization \\ and evolution of Off-forward 
parton distributions}
   
  We now turn to the infrared divergences present in all 
of the amplitudes. These divergences arise from 
the regions of loop-mementum integration where some of 
the internal propagators are near their ``mass shells". 
In these regions, perturbative calculations 
are clearly meaningless. The standard procedure of fixing 
this problem is to factorize the amplitudes into the infrared
safe (i.e. devoid of infrared divergences) and infrared
divergent pieces, interpreting the latter as nonperturbative 
QCD quanities. Of course, the factorization procedure  
has a large degree of arbitrariness. To factorize in a 
physically interesting way, one usually chooses 
the nonperturbative objects as the
parton distributions
in the target, defined in a particular renormalization 
scheme. In this paper, we consider parton distributions in the 
$\overline{\rm MS}$ scheme. The goal of this section 
is to show that all infrared divergences present 
in the Compton amplitudes can be associated with these 
distributions.

Since we consider factorization within the framework of perturbation 
theory, we need to compute the parton distributions in quark and gluon 
targets in perturbative QCD. 
At the leading order in $\alpha_s$, one has
\begin{equation}
      F_{q'/q}^0(x, \xi) = \delta_{qq'}\delta (x-1) \ ,
\end{equation}
in an onshell quark $q$. 
At next-to-leading order, one can calculate 
directly from the definitions in section II (with
the external hadron states replaced with perturbative
quark states):
\begin{eqnarray} 
  F^1_{q'/q}(x,\xi) &=& 
{\alpha_s(\mu)\over 2\pi}\delta_{qq'}
\left({2\over \epsilon} + \ln (\mu^2 e^{\gamma_E} /4\pi\mu_0^2)\right)
    \left(P(x, \xi) + A\delta(x-1) \right)\ , 
\label{F1}
\end{eqnarray}
where 
\begin{equation}
A = C_F\left\lbrack{3\over 2} +  \int^x_\xi{dy\over y-x-i\epsilon}
         +  \int^x_{-\xi}{dy\over y-x-i\epsilon}\right\rbrack \ , 
\end{equation}
and 
\begin{eqnarray}
       P(x, \xi) &=& C_F{x^2+1-2\xi^2\over (1-x+i\epsilon)(1-\xi^2)} \ ,
       ~~~~~~~~~~   x >\xi   \nonumber    \\
                 &=& C_F{x+\xi\over 2\xi(1+\xi)}
   \left(1+{2\xi\over 1-x+i\epsilon}\right)\ , ~~~~~ -\xi<x<\xi \nonumber \\
                 &=& 0 \ , ~~~~~~~~~~~~~~~~ x<-\xi \ \ .
\end{eqnarray}
The quark distributions
contain infrared divergences signaled by the presence of 
the $1/\epsilon$ terms. These divergences reflect the soft 
physics intrinsic to the parton distributions. 
$F^1_{q'/q}$, calculated for the quark 
target, satisfies the evolution equation derived in Ref. \cite{ji2} :
\begin{equation}
      {D_Q\, F_a(x,\xi)\over D\, \ln Q^2}={\alpha_s(Q^2)\over 2\pi}
\int_x^1 {dy\over y}P_{ab}\left({x\over y},{\xi\over y}\right)F_b(y,\xi),
\label{alt}
\end{equation}
where $b$ is summed over all parton species and the $P$'s are the 
off-forward Alterelli-Parisi kernels, or splitting functions.  
The `covariant' derivative is defined to include $A$, the 
endpoint contribution in Eq. (\ref{F1}).

We can reexpress the symmetric part of the quark Compton amplitude
in terms of the unpolarized, off-forward quark distribution 
\begin{eqnarray}
      T^{(ij)}_q &=& -g^{ij}\sum_{q'}e_{q'}^2\int^1_{-1} 
dx F_{q'/q}(x,\xi)\left\lbrace 
{1\over x_B-x}-{\alpha_s C_F\over 4\pi}\left\lbrace {9\over {x_B-x}}
\right.\right.\nonumber\\ &&
\left.\left.-{x\over\xi}\left\lbrack\left({2\xi\over x_B^2-x^2}
+{x_B+\xi\over x_+x_-}\right)\left(3-\ln\,
\left(1-{\xi\over x_B}\right)\right)-3{x_B-\xi\over x_+x_-}
\right\rbrack \ln\left(1-{\xi\over x_B}\right)\right.\right.\label{qu}\\
&& +\left.\left.\left\lbrack\left({x_B+x\over x_+x_-}+{2\over x_B-x}
\right)\left(3-\ln\,\left(1-{x\over x_B}\right)
\right)-3{x_B-x\over x_+x_-}-{3\over x_B-x}\right\rbrack \ln
\left(1-{x\over x_B}\right)\right\rbrace\right.\nonumber\\ &&
\left.+(x_B\rightarrow -x_B)\right\rbrace\ . \nonumber
\end{eqnarray}
Analogously, we find that the antisymmetric 
part of the quark Compton amplitude
can be expressed in terms of the polarized off-forward quark 
distribution
\begin{eqnarray}
      T^{[ij]}_q &=& i\epsilon^{\alpha\beta ij}n_\alpha p_\beta
\sum_{q'}{e_{q'}}^2 
\int^1_{-1} dx\tilde F_{q'/q}(x,\xi)\left\lbrace {1\over x_B-x}-{\alpha_s 
C_F\over 4\pi}\left\lbrace {9\over x_B-x}\right.\right.  \nonumber\\ &&
\left.\left.-\left\lbrack\left({2x_B\over x_B^2-x^2}+{x_B+\xi\over 
x_+x_-}\right)\left(3-\ln\,\left(1-{\xi\over x_B}
\right)\right)-{x_B-\xi\over x_+x_-}\right\rbrack \ln\left(1-{\xi
\over x_B}\right)\right.\right.\label{qp}\\ &&
+\left.\left.\left\lbrack\left({x_B+x\over x_+x_-}+{2\over x_B-x}
\right)\left(3-\ln\,\left(1-{x\over x_B}\right)
\right)-{x_B-x\over x_+x_-}-{3\over x_B-x}\right\rbrack \ln\left(1
-{x\over x_B}\right)\right\rbrace\right.\nonumber\\ &&
\left.-(x_B\rightarrow -x_B)\right\rbrace \ . \nonumber
\end{eqnarray}

We now turn to Compton scattering on a gluon ``target". Infrared
divergent contributions come from the regions where the quarks 
in the box diagrams are nearly onshell. Therefore, it is natural to 
associate these divergences with the quark  
distributions in a gluon target. On the other hand, the finite 
contributions come from regions where large momenta run
through the quark loop.  In these regions, the photon has an effective 
pointlike coupling with the gluons in the 
target. At leading order, the off-forward 
gluon distribution in a gluon target is just 
\begin{equation}
       F_{g/g}^0(x, \xi) = {1\over 2}(\delta(x-1)-\delta(x+1)) \ . 
\end{equation}
To order $\alpha_s$, there are quark partons in the gluon. 
The corresponding off-forward quark distribution is 
\begin{equation}
     F^1_{q/g}(x,\xi) = {\alpha_s(\mu)\over 2\pi}
\left({2\over \epsilon} + \ln (\mu^2 e^{\gamma_E} /4\pi\mu_0^2)\right)
    P(x, \xi) \ , 
\end{equation}
where for $x>\xi$
\begin{equation}
       P(x, \xi) = 2T_F {x^2+(1-x)^2-\xi^2 \over 
        (1-\xi^2)^2}\,\, ,
\end{equation}
and for $-\xi<x<\xi$
\begin{equation}
       P(x, \xi) = T_F {(x+\xi)(1-2x+\xi)\over 
              \xi(1+\xi)(1-\xi^2)}\,\, .
\end{equation}
$F^1_{q/g}(x, \xi)=0$ for $x<-\xi$. 

With the above off-forward distributions, we reexpress
the symmetric part of the gluon Compton scattering amplitude
\begin{eqnarray}
       T_g^{(ij)} &=& -{\alpha_sT_F\over 2\pi}g^{ij}\left(\sum_{q}
e_q^2\right)
\int^1_{-1} dx {x\over x_+x_-}F_{g/g}(x,\xi)\nonumber\\ &&
\left\lbrace\left\lbrack\left(1+2x_B{x_B-x\over x_+x_-}\right)\left(
4-\ln\,\left(1-{x\over x_B}\right)\right)-2\right
\rbrack \ln\left(1-{x\over x_B}\right)\right.\label{gu}\\ &&
-\left.2x_B\left\lbrack\left({x_B\over x_+x_-}-{1\over 2\xi}\right)
\left(4-\ln\,\left(1-{\xi\over x_B}\right)\right)+
{1\over\xi}\right\rbrack\left(1-{\xi\over x_B}\right)\ln\left(1-
{\xi\over x_B}\right)\right\rbrace\nonumber\\ &&
- g^{ij}\sum_{q}{e_{q}^2}\int^1_{-1} dx F_{q/g}(x, \xi)
     {1\over x_B-x} \nonumber \\ &&
 +(x_B\rightarrow -x_B)\nonumber \ . 
\end{eqnarray}          
Similarly, we can reexpress the antisymmetric part of the gluon
Compton amplitude in terms of helicity-dependent, off-forward
quark and gluon distributions
\begin{eqnarray}
      T_g^{[ij]} &= &i{\alpha_sT_F\over 2\pi}\epsilon^{\alpha\beta ij}
n_\alpha p_\beta\left(\sum_{q}e_q^2\right)
\int^1_{-1} dx {x\over x_+x_-}\tilde F_{g/g}(x,\xi)\nonumber\\ &&
\left\lbrace\left\lbrack\left(1+2x{x_B-x\over x_+x_-}\right)\left(
4-\ln\,\left(1-{x\over x_B}\right)\right)-2\right
\rbrack \ln\left(1-{x\over x_B}\right)\right.\label{gp}\\ &&
-\left.2x_B\left\lbrack{x\over x_+x_-}\left(4-\ln\,
\left(1-{\xi\over x_B}\right)\right)\right\rbrack\left(1-{\xi\over
 x_B}\right)\ln\left(1-{\xi\over x_B}\right)\right\rbrace\nonumber\\ &&
+i\epsilon^{\alpha\beta ij}n_\alpha p_\beta\sum_{q}{e_{q}^2}
\int^1_{-1} dx \tilde F_{q/g}(x, \xi)
     {1\over x_B-x} \nonumber \\     &&
- (x_B\rightarrow -x_B) \nonumber
\end{eqnarray}        

We summarize the one-loop Compton amplitude on a target $N$
 in the factorization formula
\begin{eqnarray}
   T^{ij}_N &=& -g^{ij} \int^1_{-1} {dx\over x} \left[\sum_{q} 
F_{q/N}(x, \xi) C_q\left({x\over x_B}, {\xi\over x_B}\right)
+F_{g/N}(x, \xi) C_g\left({x\over x_B}, 
{\xi\over x_B}\right) \right] \nonumber \\
&& +  i\epsilon^{ij\alpha\beta}n_\alpha p_\beta 
\int^1_{-1} {dx\over x} \left[\sum_{q}
\tilde F_{q/N}(x, \xi) \tilde C_q\left({x\over x_B}, {\xi\over x_B}\right)
+\tilde F_{g/N}(x, \xi) 
\tilde C_g\left({x\over x_B}, {\xi\over x_B}\right) \right]\,\,\, ,
\label{fac}
\end{eqnarray}
where $C_q$, $C_g$, $\tilde C_q$ and $\tilde C_g$ are shown to
order $\alpha_s$ in Eqs. (\ref{qu}, \ref{gu}, \ref{qp}, \ref{gp}), 
respectively.  [We emphasize here again that we have neglected
the contributions from longitudinally polarized photons and 
from photon helicity flip. Both effect start at 
order $\alpha_s$.]
In the above form, all infrared sensitive contributions
have been isolated in the relevant parton distributions, 
which must be calculated nonperturbatively or 
measured in experiments.  In the DVCS limit
$\xi\rightarrow x_B$, the coefficient functions
remain finite, although they have branch cuts there.
This indicates that factorization holds 
for two-photon amplitudes even when one of the photons
is onshell. We will argue in the next section 
that the above formula, one of the
main results of this paper, remains valid to 
all orders in perturbation theory. 
     
\section{Factorization of DVCS amplitudes to all orders}

In this section, we generalize the one-loop result of
the previous section, showing that the 
factorization formula Eq. (\ref{fac}) 
is valid in the DVCS limit
to all orders in perturbation theory.  
The one-loop result indicates
that all soft divergences - those associated with
integration regions where all components of some internal 
momenta are zero - cancel,  
whereas all collinear divergences can be factorized 
into the off-forward parton distributions. To see that 
this happens also at higher orders in perturbation theory, 
it is important to understand how the soft cancellation 
happens in the simplest case.

The self-energy diagram in Fig. 2b does not contain any infrared 
divergences because the intermediate quark is far offshell. 
The vertex corrections in Figs. 2c and 2d potentially 
have infrared divergences, but a simple power counting 
indicates that these diagrams are in fact infrared convergent. 
Thus infared divergences appear only in the box diagram. 
In the region where the gluon momentum $k$ is soft, we can 
approximate the integral as 
\begin{equation}
      \sim  \int {d^4k\over (2\pi)^4} {p'\cdot p}
       {1\over p'\cdot k} {1\over p\cdot k} {1\over k^2} \ , 
\end{equation}
where $p$ and $p'$ are the momenta of the two external
quark lines.
On the other hand, the wave function renormalization
of an ``on-shell'' quark, $\delta Z$,  also contains infrared divergences. 
Grouping these divergent terms together, we have the entire
soft contribution 
\begin{equation}
   \sim   -{1\over 2}\int {d^4k\over (2\pi)^4}\left({p^\mu\over p\cdot k}
      - {p'^\mu\over p'\cdot k}\right)^2 {1\over k^2} \ . 
\end{equation}
In the collinear approximation, $p'$ is proportional to 
$p$ and thus the above integral vanishes. 

For higher-order Feynman diagrams, a systematic method of
identifying, regrouping and factorizing infrared-sensitive 
contributions has been developed by Libby, Sterman, Collins, 
and others \cite{factor}. The method essentially consists of  
the following steps: 1) simplify the Feynman integrals by 
setting all the soft scales to zero, including the 
quark masses; 2) identify the regions of loop integration 
which give rise to infrared divergences; 3) use 
infrared power counting to find the leading 
infrared-divergent regions; 4) show
that all soft and collinear divergences either cancel 
or factorize into some nonperturbative quantities. 
In the remainder of this section, we examine the validity of
the factorization formula Eq. (\ref{fac}) in the limit
of $x_B=\xi$ following the above steps. 

In DVCS, the leading contributions come essentially
from a massless collinear process in which the external momenta
take the form shown in Eq. (6). In this simplified 
kinematic region, all infrared-sensitive contributions
appear as $1/\epsilon$ poles in dimensional regularization. 
If a contribution contains no infrared divergences, 
it comes from regions of loop momenta comparable to 
the hard scale $Q^2$, and thus it is insensitive to the 
soft scales. An infrared divergent contribution
must come from the integration regions 
where some internal propagators are near their mass shells. 
Since such soft contributions cannot be calculated 
reliably in perturbative QCD and eventually must be 
taken into account with nonperturbative 
matrix elements, one can use any valid infrared regulator 
to characterize them in perturbative calculations. 
Thus the collinear massless limit helps
to simplify the identification of soft contributions
while leaving the truly-perturbative contributions intact. 

Infrared divergences appear
in a Feynman diagram when some of the external 
momenta are onshell. The regions of integration
producing such contributions can be identified from 
the Landau equations which are derived by considering
the analytical properties of the diagrams as functions
of complex external momenta.  
According to Coleman and Norton \cite{coleman}, these regions 
can be represented by the so-called reduced diagrams in which 
offshell lines are shrunk to points and onshell lines
are drawn according to their real space-time propagation.
We shall argue below that the general {\it leading} reduced 
diagram for DVCS is the one shown in Fig. 5a, in which an 
incoming virtual photon 
and an outgoing real one are attached to the hard 
interaction blob, which in turn is connected
to the forward nucleon jet with two 
collinear quark lines or two physically polarized 
gluon lines, plus an arbitrary number of longitudinally-polarized 
collinear gluon lines. 

\begin{figure}
\label{fig5}
\epsfig{figure=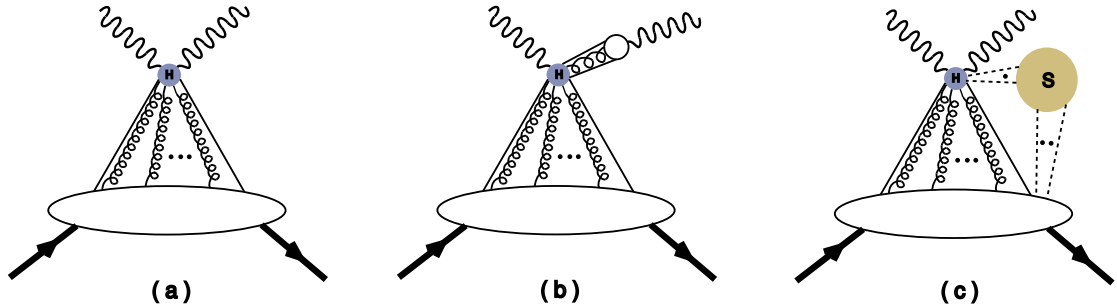,height=4cm}
\caption{General reduced diagrams for the DVCS process.}
\end{figure}           
To decide that a reduced diagram is leading, one can
use infrared power counting \cite{sterman}, 
which is essentially a light-cone dimensional 
analysis \cite{collins}. A simple way to 
proceed is to consider the mass dimensions of 
the soft vertices that connect lines with either collinear 
or soft momenta. Since the dimension of an amplitude is 
fixed, all soft mass dimensions must be 
compensated by the hard scale $Q^2$. 
Assuming covariant normalization for the
external states ($\langle p|p\rangle = 2p^0(2\pi)^3
\delta^3(0)$), every external wave function
contributes mass dimension $-1$. The collinear quarks 
and gluons into a soft hadron vertex have 
effective mass dimensions depending on their
polarizations. A Dirac field $\psi$ can be written
as a sum of good ($\psi_+$) and bad ($\psi_-$)
components, where $\psi_{\pm} = P_{\pm}\psi$ and
$P_{\pm} = {1\over 2}\gamma^{\mp}\gamma^{\pm}$. 
The good (bad) component has effective 
light-cone mass dimension $1$ ($2$). A vector 
potential $A^\mu$ has light-cone
components $A^+$, $A^\perp$, and $A^-$, which have
effective mass dimensions 0, 1, and 2, respectively. 
For the reduced diagram shown in Fig. 5a, the only 
soft mass dimension comes from the 
nucleon-quark-gluon blob. Using the above 
rule, we find it is 0 [$0=2$(physical parton lines $-2$(
external nucleon states)]. Because $T^{\mu\nu}$ is
dimensionless, the leading reduced diagram contributes at order 
${\cal O}(Q^0)$. 

It is somewhat surprising that the leading 
region is independent of the virtuality of the 
final state photon as long as the initial photon
is deeply virtual. When the final state photon is
real, it can have pointlike coupling to quarks as well as 
extended coupling via its soft wave function, 
as happens in the case of vector dominance. Thus one 
has an additional reduced diagram in which a jet 
of quarks and gluons emerges in the $n^\mu$ direction
and combines into a real photon long after
the hard scattering (Fig 5b). Such a reduced 
diagram has already been considered in 
Ref. \cite{collins} and is ${\cal O}(Q^{-1})$ by
infrared power counting. Indeed, according to the
discussion in the previous paragraph, the photon
wave function vertex has a soft mass 
dimension $1 = 2$ (quark lines) $- 1$ (photon state)
(recall the dimensionful pion decay constant $f_\pi$). 
A negative hard power $(Q^{-1})$ is needed 
in $T^{\mu\nu}$ to balance it out. 

The power counting involving soft quark and gluon lines
is more subtle, and some discussion may be found 
in Ref. \cite{collins}. The result is that any 
reduced diagram with soft lines connecting 
the hard scattering blob to the nucleon jet 
is subleading (Fig. 5c). The
situation here is exactly analogous to the case 
of forward virtual Compton scattering relevant to 
deep-inelastic scattering as discussed,
for instance, by Sterman \cite{sterman}. 
A simple example is the vertex correction 
diagram we already discussed
above. When the gluon becomes soft, it has a reduced 
diagram like Fig. 5c. However, it has no contribution
at leading power. 

We now come back to the leading reduced diagram
shown in Fig. 5a. The collinear gluons with longitudinal 
polarization connect the hard scattering part and
the nucleon jets. These collinear gluons can be factorized
using the generalized Ward identities as described in 
Ref. \cite{factor}.  Eventually, all 
collinear gluons can effectively be attached
to an eikonal line in the light-cone direction 
conjuate to the nucleon jet, $n^\mu$. Physically, this 
means that as far as the collinear gluons 
are concerned, the hard interaction part acts as a 
jet of particles propagating along $n^\mu$. 
The internal structure of the hard interaction cannot be 
resolved and thus only the total color charge
and momentum of the jet is relevant. The eikonal line 
together with the physical quarks and gluons and 
the nucleon jets form the off-forward parton
distributions defined in Section II. 
In the hard scattering, only the total momentum and 
charge supplied by collinear partons are important.
Thus, one can calculate it with 
incoming physical partons carrying the total momentum
and charge of all the collinear longitudinally-polarized
gluons. In this way, we have a complete 
factorization of the soft and hard physics
in the DVCS process.

\section{Generalized Operator Product Expansion \\ 
and Wilson's Coefficients to the NLO order}

     The factorization formula in Eq. (\ref{fac}) 
for the general Compton scattering process can 
also be examined in the form of an operator product expansion. 
The OPE was first introduced by Wilson \cite{wilson} 
in 1969 and has been used extensively in deep inelastic
scattering and other perturbative QCD processes.
For the product of two currents
separated near the light-cone, the expansion is
threefold. Primarily, it is a twist expansion, 
in which twist-two contributions are leading whereas
the higher twist terms are suppressed by
powers of $1/Q^2$.  Each term in the twist expansion
contains an infinite number of local operators of the 
relevant twist. This may be thought of as a kind of
Taylor expansion of bilocal operators along the 
light-cone. Finally, the coefficients of local operators 
(Wilson coefficients) are themselves expansions in the strong coupling constant.  
The Wilson coefficients for the unpolarized
DIS process were calculated at order $\alpha_s$ in the $\overline{\rm MS}$
scheme in \cite{bardeen}.  For the polarized 
case, one can find them in \cite{pol1}.  

When considering off-forward processes, the expansion 
of an operator product must include operators with total
derivatives. We call this expansion the generalized OPE.
In the remainder of this section, we
will recast our factorization formula in its generalized
OPE form. In the process, we identify these total derivative 
operators and obtain their Wilson coefficients 
to next-to-leading order in $\alpha_s$. 
The final result agrees with the known OPE 
in the DIS limit ($\xi\rightarrow 0$).  

To derive the generalized OPE, 
we expand $T^{ij}$ in a power 
series about $x_B=\infty$.  In this way, we 
can express the amplitude in terms of
moments of the parton disributions rather than 
the distributions themselves.  Eventually, we will 
relate the moments of parton distributions
to the matrix elements of local operators.
After the aforementioned expansion, we have
\begin{eqnarray}
     && T^{ij}=-g^{ij}T_S+i
\epsilon^{\alpha\beta ij}n_\alpha p_\beta
T_A\,\,\, ;\nonumber\\
     && T_S = 2 \sum_{n\, \rm even=2}^{\infty}\,\,\sum_{m\, 
\rm even=0}^{\infty}\int^1_{-1}{dx\over x}{\left({x\over x_B}
\right)}^n{\left({\xi\over x_B}\right)}^m\left\lbrack 
\sum_q c^q_{nm}F_q(x,\xi)+c^g_{nm}F_g(x,\xi)\right\rbrack
\nonumber \\
     && T_A = 2 \sum_{n\, \rm odd=1}^{\infty}\,\,\sum_{m\, 
\rm even=0}^{\infty}\int^1_{-1}{dx\over x}\left({x\over x_B}\right)^n
\left({\xi\over x_B}\right)^m\left\lbrack \sum_q \tilde c^q_{nm}
\tilde F_q(x,\xi)+\tilde c^g_{nm}\tilde F_g(x,\xi)\right \rbrack, 
\label{OPE}
\end{eqnarray} 
The coefficients for the moments
of the quark distributions in the expansion are
\begin{eqnarray}
     c_{nm}^q &=& \delta_{m0}-{\alpha_sC_F\over 4\pi}\left\lbrace
\left\lbrack 9-{8\over n}+{2\over n+1}+4S_2(n-1)-4T_1^1(n-1)\right.
\right.\nonumber \\ 
&& \left. -S_1(n-1)\left(3+{2\over n}+{2\over n+1}\right)\right]
\delta_{m0} + \left[ {6n\over m(m+n)}-{1\over(n+m)(n+m+1)}\right.
 \nonumber \\ && \left.\left.
+ {4\over m}S_1(m-1)-2S_1(n+m-1)\left({1\over n+m}+{1\over n+m+1}
\right)\right](1-\delta_{m0})\right\}\ , \nonumber\\
\\
    \tilde c_{nm}^q &=& \delta_{m0}-{\alpha_sC_F\over 4\pi}\left
\lbrace\left\lbrack 9-{6\over n}+4S_2(n-1)-4T_1^1(n-1) 
     \right.\right.\nonumber \\
   && \left. -S_1(n-1)\left(3+{2\over n}+{2\over n+1}
     \right)\right]\delta_{m0}
   +\left[ {6n\over m(n+m)}-{3\over m}+{4\over m}S_1(m-1)\right.
    \nonumber \\ 
    && \left.\left.
      -2S_1(n+m-1)\left({1\over n+m}+{1\over n+m+1}\right)\right]
    \left(1-\delta_{m0}\right)\right\} \ ,   
\end{eqnarray}
where we have introduced
\begin{eqnarray}
     && S_j(n):=\sum_{i=1}^n{1\over i^j} \ , \nonumber\\
     && T_j^k(n):=\sum_{i=1}^n{S_j(i)\over i^k} \ . 
\end{eqnarray}
Notice that the above expansion contains only positive powers
of $x$ and $\xi$. This result
is not immediately obvious because of the $x_+x_-$ denominators 
in the amplitudes. In the case of the gluon distribution 
functions, we have an additional factor of $x_+x_-$ in 
the denominator. Since the final OPE contains only local operators,
these factors have to be cancelled in the process of 
expansion. Indeed, this turns out to be the case 
and we obtain the coefficients of the positive
moments of the gluon distributions
\begin{eqnarray}
      c_{nm}^g &=&{\alpha_sT_F\over 2\pi}\left\lbrack 
{m\over n+m}-{m+2\over n+m+2}-2S_1(n+m-1)\left({1\over n+m}
-{m+2\over n+m+1}+{m+2\over n+m+2}\right)\right\rbrack\nonumber\\
\\
      \tilde c_{nm}^g  &=&{\alpha_sT_F\over 2\pi}\left
\lbrack 2+2S_1(n+m-1)\right\rbrack\left({m+1\over n+m}-
{m+2\over n+m+1}\right)\left(1-\delta_{n1}\right)\ . \nonumber
\end{eqnarray}

Having obtained an expansion involving the moments of the 
distributions, we move toward a general form of
the OPE.  To this end, we consider the moments 
of the parton distributions.  We begin by 
observing that for quarks
\begin{equation}
     \int^1_{-1} dx x^{n-1}F_q(x,\xi)={1\over 2}
\left\langle P_f\left|\bar\psi_q(0)\stackrel{\leftrightarrow}
{i\partial}_{\mu_1}
\cdots\stackrel\leftrightarrow{i\partial}_{\mu_{n-1}}
\gamma_{\mu_n}\psi_q(0)\right|P_i\right\rangle n^{\mu_1}\cdots n^{\mu_n}
\end{equation}
holds in light-cone gauge, where we have defined $\stackrel
\leftrightarrow\partial\,={1\over 2}(\stackrel\rightarrow
\partial-\stackrel\leftarrow\partial)$.  The parton distribution
depends on $x$ only through the exponential, 
which allows one to integrate over the $x$ through simple
partial integrations. The $\lambda$ integration 
can then be done trivially. So the moments of the quark
distribution can be expressed in terms of the matrix elements of
local operators in the light-cone 
gauge.  However, in this gauge, these gauge-dependent 
local operators are equal to the gauge invariant operators obtained 
by replacing the partial 
derivatives in the above expression by covariant derivatives, 
\begin{equation}
      _q{\cal O}_{\mu_1\mu_2\cdots\mu_n}^n=\bar 
\psi(0)\stackrel\leftrightarrow{i\cal D}_{(\mu_1}\cdots \stackrel
\leftrightarrow{i\cal D}_{\mu_{n-1}}\gamma_{\mu_n)}\psi(0),
\end{equation}
where $(\cdots )$ signifies that 
the indices are symmetrized and the trace has been removed. 
Thus the moments of the quark distribution functions are just matrix 
elements of the + components of the above operators between 
the initial and final 
hadron states.  We also recognize that 
\begin{equation}
      (n\cdot i\partial)\,\, _q{\cal O}_n^{+\cdots +} 
  = 2\xi{\cal O}_n^{+\cdots +} \ .
\end{equation}
This prompts us to define
\begin{eqnarray}
      _q{\cal O}^{n,m}_{\mu_1\mu_2\cdots\mu_n} & = & i\partial_{\,(\mu_1}
\cdots i\partial_{\,\mu_m}\bar\psi(0)\stackrel\leftrightarrow
{i\cal D}_{\mu_{m+1}}\cdots\stackrel\leftrightarrow{i\cal 
D}_{\mu_{n-1}}\gamma_{\mu_n)}\psi(0)\ , \nonumber \\ 
      _q{\tilde {\cal O}}^{n,m}_{\mu_1\mu_2\cdots\mu_n} & = & 
i\partial_{\,(\mu_1}
\cdots i\partial_{\,\mu_m}\bar\psi(0)\stackrel\leftrightarrow
{i\cal D}_{\mu_{m+1}}\cdots\stackrel\leftrightarrow{i\cal 
D}_{\mu_{n-1}}\gamma_{\mu_n)}\gamma_5\psi(0)\ . \nonumber \\
     _g{\cal O}^{n,m}_{\mu_1\mu_2\cdots\mu_n} & = &i\partial_{\,(\mu_1}
\cdots i\partial_{\,\mu_m}F_{\mu_{m+1}\alpha}(0)\stackrel\leftrightarrow
{i\cal D}_{\mu_{m+2}}\cdots\stackrel\leftrightarrow{i\cal
D}_{\mu_{n-1}}F^\alpha_{~\mu_n}(0)\ , \nonumber \\ 
     _g{\tilde {\cal O}}^{n,m}_{\mu_1\mu_2\cdots\mu_n} & 
   = &i\partial_{\,(\mu_1}
\cdots i\partial_{\,\mu_m}F_{\mu_{m+1}\alpha}(0)\stackrel\leftrightarrow
{i\cal D}_{\mu_{m+2}}\cdots\stackrel\leftrightarrow{i\cal
D}_{\mu_{n-1}}i\tilde F^\alpha_{~\mu_n}(0)\ . 
\end{eqnarray}
After replacing the moments of parton distributions in Eq.(\ref{OPE})
with matrix elements of these operators and 
interpreting the result as an operator relation, we find the following
generalized OPE,
\begin{eqnarray}
     &&~~~ i\int d^{\,4}z~e^{iq\cdot z}~
    TJ^{\, \nu}\left({z\over 2}\right)J^{\, \mu} \left(-{z\over 2}\right)
   \nonumber \\ 
&=& (-g^{\mu\nu}+\cdots )\sum_{n~\rm even =2}^{\infty}\,\,
\sum_{m~\rm even=0}^{n}
\left({2^{n-m}q_{\mu_1}\cdots q_{\mu_n}\over (Q^2)^n}\right)
 \sum_{a=q,g} c^a_{n-m,m}\,\,_a{\cal O}_{n,m}^{\mu_1
\cdots \mu_n}  \nonumber \\ 
&& + i\epsilon_{\mu\nu \alpha\beta} q^\alpha 
\sum_{n~\rm odd = 1}^{\infty}\,\,\sum_{m~\rm even=0}^{n}
\left({2^{n-m}q_{\mu_2}\cdots q_{\mu_n}\over 
(Q^2)^n}\right)  \sum_{a=q,g} \tilde c^a_{n-m,m}\,
\,_a\tilde{\cal O}_{n,m}^{\beta \mu_2\cdots \mu_n} + ...
\label{ope2}
\end{eqnarray}
It must be pointed out that the generalized OPE does not 
have a unique form. One can define $x_B$ as any dimensionless 
invariant formed from the external momenta which remains finite
in the Bjorken limit and expand the amplitude in inverse powers
of this variable.  This will lead to
a different set of coefficient functions, but the physical 
content is the same.  The choice of which OPE to use is determined
by the specifics of the problem at hand.

Of course, the above expression contains only the contributions
to leading order in $1/Q^2$. Since the operators ${\cal O}_{n,m}$ are 
symmetrized and traceless, their rank is $n$.  
The mass dimension of these operators is just the 
dimension of the fermion fields plus that of the derivatives, or $3+n-1$.  
Hence these operators are all twist 2. At the next order in $1/Q^2$
one has to consider operators of higher twist, which are beyond
the scope of this paper. 

\section{summary and comments}

In this paper, we have studied the QCD factorization for 
deeply virtual Compton scattering explicitly at one loop and
then to all orders in perturbation theory. Our conclusion 
is that DVCS is factorizable in perturbation theory. 
This statement has the same level of rigor as 
the ordinary operator production expansion 
used in deep-inelastic scattering. In fact, assuming
the generalized OPE with total derivative operators, 
DVCS can be recovered by analytically continue
$x_B$ variable from $x_B>1$ region to the point $x_B=\xi$. The 
factorization theorem guarantees that the Compton
amplitudes are finite there, although the one-loop 
calculation indicates that they are not analytic. 

We have also computed the coefficient functions 
to order $\alpha_s$ for the generalized OPE including the 
total derivative operators. For general 
two photon processes, one has to include 
the longitudinal photon scattering, which has been done
in Ref. \cite{man}, and photon-helictity flip 
amplitude \cite{hoodbhoy}. The scale 
evolution of total derivative operators can best be
studied using conformally-symmetric operators \cite{muller}. 
In fact, it has been known for a long time that 
at the leading-log level, the operators of same twist and
dimension evolve multiplicatively in Gegenbauer 
polynomial combinations. It is a simple excercise to 
transform Eq. (\ref{ope2}) into this basis. 

Note Added: After this work was completed, we learned 
that the DVCS factorization has also been studied 
by Collins and Freud \cite{freud}.  Their arguments
and conclusions are similar to ours. 
We also learned that some 
aspects of higher-order corrections to DVCS have been  
considered by Belitsky and Sch\"afer \cite{bel}. 

\acknowledgements
We thank J. Collins, A. Mueller, A. Radyushkin, 
and G. Sterman for conversations about factorization. 
This work is supported in part by funds provided by the
U.S.  Department of Energy (D.O.E.) under cooperative agreement
DOE-FG02-93ER-40762.

\end{document}